\documentclass[onecolumn]{IEEEtran}
\usepackage{cite}
\usepackage{amsmath,amssymb,amsfonts}
\usepackage{algorithmic}
\usepackage{graphicx}
\usepackage{textcomp}
\usepackage{color}
\newtheorem{theo}{Theorem}

\newtheorem{obs}{Remark}

\newtheorem{propo}{Proposition}

\newtheorem{example}{Example}
\newtheorem{assum}{Assumption}
\newtheorem{proof}{Proof}

\DeclareMathOperator{\rank}{rank}

\def\BibTeX{{\rm B\kern-.05em{\sc i\kern-.025em b}\kern-.08em
    T\kern-.1667em\lower.7ex\hbox{E}\kern-.125emX}}
\begin{document}
\title{Super Twisting based  Lyapunov Redesign  for Uncertain Linear Delay Systems}
\author{Marco A. Gomez,  Christopher D.  Cruz-Ancona,  and Leonid Fridman, \IEEEmembership{Member, IEEE}
\thanks{This research was supported by DGAPA-UNAM (Programa de Becas Posdoctorales DGAPA en la UNAM), CONACyT (Consejo Nacional de Ciencia y Tecnolog \'ia), Project 282013, and CVU 833748; PAPIIT–UNAM (Programa de Apoyo a Proyectos de Investigaci\'on e Innovaci\'on Tecnol\'ogica) IN115419 }
\thanks{M. A. Gomez is with the Department of Mechanical Engineering, DICIS,  Universidad de Guanajuato, 36885 Slamanca, Gto., Mexico (e-mail: marco.gomez@ugto.mx). }
\thanks{C.D. Cruz-Ancona and L. Fridman are with the Department of Robotics and Control, Engineering Faculty, Universidad Nacional Aut\'onoma de M\'exico, 04510, Mexico (e-mail: cdiegoca89@gmail.com; lfridman@unam.mx).}
}

\maketitle

\begin{abstract}
We present a new continuous Lyapunov Redesign (LR)  methodology for the robust stabilization of  a class of uncertain time-delay systems that  is based on the so-called Super Twisting Algorithm. The main feature of the proposed approach is that allows one to \textit{simultaneously} adjust the chattering effect  and achieve asymptotic stabilization of the uncertain system, which is lost  when continuous approximation  of the unit control is considered. At the basis of the Super Twisting based LR methodology is a class of Lyapunov-Krasovskii functionals, {whose} particular form of its time derivative  allows one  to define a delay-free sliding manifold on which some class of smooth uncertainties are compensated. 
\end{abstract}


\section{Problem statement}  
We  consider uncertain linear time-delay system of the form
\begin{equation}
\begin{split}
\label{ec:linear_sys}
\dot x (t)&=\sum_{j=0}^{m}A_jx(t-h_j)+B(u(t)+\delta(t,\bar x)),\:t\geq 0,\\
x(t)&=\varphi(t),\: t\in [-h_{m},0],
\end{split}
\end{equation}
where the initial function $\varphi$ is considered to belong to the space of   continuous functions, $x(t)\in \mathbb{R}^n$ is the state at present time, $A_j\in \mathbb{R}^{n\times n}$, $j=0,\ldots,m$,  $B\in \mathbb{R}^{n\times k}$,  the delays $0=h_0<h_1<\ldots <h_{m}$ are known, and the vector $\bar x^T(t):=\begin{pmatrix}
x^T(t) & x^T(t-h_1) & \ldots & x^T(t-h_m)
\end{pmatrix}$.  The delayed state dependent uncertainty $\delta(t,\bar{x})$ is continuous in $\bar{x}$, for all $t\in \mathbb{R}$,  and it is Lebesgue measurable in $t$,  for all $\bar{x}\in \mathbb{R}^{(m+1)n}$. {As specified by Assumption \ref{ass:delta} in Section \ref{sec:STA},  it is assumed that it can be divided into vanishing and non-vanishing terms.  The non-vanishing terms in $\delta$ are assumed to be continuously differentiable in time. Moreover, its derivative  and the vanishing terms are bounded by time dependent Lebesgue integrable functions (cf.  with Chapter 3 in \cite{Kolmanovskii1999})}.

Systems of the form \eqref{ec:linear_sys} are pervasive in engineering \cite{Kolmanovskii1999}, and design of robust  control algorithms  that  mitigate the  perturbation effects  has been object of active research in the last  decades { \cite{Richard2001, Han2009,Han2010, Oliveiraetal2016, Sanchezetal2020}}. As in systems without delays, Lyapunov redesign (LR) \cite{Gutman1979,Leitmann1979,Khalil1992} methodology can be used to design a robust control law to stabilize time delay system \eqref{ec:linear_sys} as long as two  requirements are fulfilled {\cite{ Thowsen1983, Wu2004, Wu2009, Rodriguez2015etal, Rodriguezetal2019}}. The first one is the existence of  a nominal control law that stabilizes the nominal system. This is, for system \eqref{ec:linear_sys} with $\delta(t, \bar x)=0$, there exist $K_j\in \mathbb{R}^{k\times n},\:j=0,\ldots,m,$ such that
\begin{equation}
\label{ec:control_nom}
u(t)=v_{nom}(t)=\sum_{j=0}^{m}K_{j}x(t-h_j)
\end{equation} renders an asymptotical stable trivial solution of the closed-loop nominal system
\begin{equation}
\label{ec:sys_cl}
\begin{split}
\dot x(t)=&\sum_{j=0}^{m}G_jx(t-h_j),\:t\geq 0,\\
x(t)=&\varphi(t),\:t\in [-h_{m},0],
\end{split}
\end{equation}
where $ G_j:=A_j+BK_{j}$ for $j=0,\ldots,m$.   The second requirement is the existence of a Lyapunov-Krasovskii functional (or Lyapunov-Razumikhin function) such that it satisfies {standard}  lower and upper bounds, and its time derivative along the solutions of the closed-loop system \eqref{ec:sys_cl} is negative. In this paper, we consider the following well-known class of Lyapunov-Krasovskii functionals   \cite{fridman2014introduction}
\begin{equation}
\begin{aligned}
\label{ec:LK_functional_gen}
V(\varphi)&=\varphi^T(0)P\varphi(0)+\sum_{j=1}^{m}\int_{-h_j}^{0}F_j\left(\vec{h},\theta,\varphi(\theta)\right) d\theta,
\end{aligned}
\end{equation}
where $P\in \mathbb{R}^{n\times n}$ is a positive definite matrix, $\vec{h}:=\begin{pmatrix}
h_1,\ldots,h_{m}
\end{pmatrix}$,  and $F_j$ are  continuous functions.

We condensate the above  requirements in the LR methodology within the following assumption, which is regarded to be satisfied from now on:
\begin{assum}
	\label{ass:func}
There exists a Lyapunov-Krasovskii functional of the form \eqref{ec:LK_functional_gen} such that for some  $\alpha_1,\:\alpha_2, \:\alpha_3\in \mathbb{R}_+$
\begin{eqnarray}
\label{ec:bound_cond}
\alpha_1 \|\varphi(0)\|^2 \leq V(\varphi)\leq \alpha_2 \|\varphi\|_{\mathcal{H}}^2\\
\label{ec:derivative_cond}
\left. \dot V(x_t)\right \rvert_{\eqref{ec:sys_cl}}\leq -\alpha_3 \|x(t)\|^2,
\end{eqnarray}
where $\Vert \cdot \Vert_{\mathcal{H}}$ is  introduced later on.
\end{assum}

\textit{Classical LR}. The LR  consists in adjusting  the control law $u(t)$ by adding a robustifying discontinuous controller $v(t)$ to the nominal one, i.e. $u(t)=v_{nom}(t)+v(t)$, such that $v(t)$ is capable of compensating the uncertainty within the time derivative of the Lyapunov-Krasovskii functional, while recovering the negative sign in its upper-bound. More specifically, notice that the time derivative of  the functional  $V$ along the solutions of uncertain system \eqref{ec:linear_sys} yields
	\begin{equation*}
	\begin{split}
	\left.\dot V(x_t)\right \rvert_{\eqref{ec:linear_sys}}\leq -\alpha_3\|x(t)\|^2  +2x^T(t)PB\left(v(t)+\delta(t,\bar x)\right).
	\end{split}
	\end{equation*} 
By taking the unit control
\begin{equation}
\label{ec:LR_discon}
v(t)=-\rho_{\delta}(t,\bar x)\frac{2B^TPx(t)}{\|2B^TPx(t)\|},
\end{equation} where $\rho_{\delta}$ is a known function such that $\|\delta(t,\bar x)\|\leq \rho_{\delta}(t,\bar x)$, 
one obtains
\begin{equation}
\label{ec:ineq_LR}
2x^T(t)PB(v(t)+\delta(t,\bar x))\leq 0.
\end{equation}
Thus,  the negativeness of the time derivative of the functional is recovered and the asymptotic stability of  system \eqref{ec:linear_sys} is ensured ({cf.} Theorem 3.1. in \cite{fridman2014introduction}). The main drawback of this approach resides in the discontinuity of the control law \eqref{ec:LR_discon} in a switching manifold  of relative degree one, which produces the undesirable chattering effect. 

\textit{Continuous LR}. In order to make a continuous LR, a continuous approximation of the unit controller \eqref{ec:LR_discon}  can be considered \cite{Ryan1984}.   Namely,
\begin{eqnarray}\label{ec:LR_approx}
v(t)\!=\!  \left \{ \begin{matrix} -\rho_{\delta}(t,\bar x)\frac{2B^TPx(t)}{\|2B^TPx(t)\|}, \; \rho_{\delta}(t,\bar x) \Vert {2B^TPx(t)} \Vert \geq \varepsilon , \\
-\rho_{\delta}^2(t,\bar x)\frac{2B^TPx(t)}{\varepsilon},\;   \rho_{\delta}(t,\bar x) \Vert {2B^TPx(t)} \Vert < \varepsilon
\end{matrix}\right.
\end{eqnarray}
where $\varepsilon$ is a given real positive number.  However, with such an approximation it is not possible to recover a  time derivative of $V$ with definite sign anymore. Indeed,  
\begin{equation*}
\begin{split}
2x^T(t)PB(v(t)+\delta(t,\bar x))
\leq -\rho_{\delta}^2 \frac{\Vert 2B^TPx(t)\Vert^2}{\varepsilon}+\Vert 2B^{T}Px(t) \Vert \rho_{\delta}\leq \frac{\varepsilon}{4},
\end{split}
\end{equation*}
and the perturbation within the time derivative of the functional is no longer compensated. The system's solutions are restricted to an arbitrarily small neighborhood of the trivial solution in a sufficiently large time and,  to maintain the solutions bounded in this region,  extremely high gains of the controller are required. Thus,  as one cannot  ensure \textit{asymptotic stability}  of the uncertain system \eqref{ec:linear_sys} anymore, the critical issue is the compromise between controller effort and the attainable residual set where the solutions will be contained. This approach was used  with  Lyapunov-Razumikhin functions in the early work \cite{Thowsen1983}, and similarly later appeared  with  Lyapunov-Krasovskii functionals in \cite{Wu2004,Wu2009} for the design of  adaptive control algorithms.  

\textit{Contribution}. An open question of practical interest within LR methodology is whether it is possible to  design a continuous controller that { \textit{simultaneously} makes system \eqref{ec:linear_sys} asymptotically stable and adjusts the chattering effect, at least for systems with fast actuators \cite{PerezFridman2019}}. In this paper, inspired by the ideas introduced in \cite{SuDrakonov1996,Estradaetal2020},  we look at  the classical LR  from a second order Sliding Mode Control (SMC) perspective by defining the sliding variable as 
\begin{equation}
\label{ec:sliding_var}
w(t):=2B^TPx(t)
\end{equation}
and the sliding manifold by
\begin{equation*}
\mathcal{S}:=\{x\in \mathbb{R}^n: w(t)=0 \}.
\end{equation*}
We propose a continuous LR methodology that relies on the super twisting algorithm (STA), a well-known technique within the SMC framework that ensures a stable second order sliding mode on the manifold $\mathcal{S}$  \cite{Levant1993,Moreno2011, MorenoOsorio2012, Vidaletal2016}. Since  the discontinuous term is integrated in the STA, the control signal is absolutely continuous at the expense of theoretically exact compensation of smooth perturbations.  By using  the STA instead of any continuous approximation of unit control, we ensure the asymptotic stability of the trivial solution of the uncertain system \eqref{ec:linear_sys}, which is the premise of a classical LR. Moreover,  STA does not  require a boundary layer strategy as in \eqref{ec:LR_approx} but a single robustifying controller does  the task. Notice that, in contrast to the classical LR where one looks for inequality \eqref{ec:ineq_LR} to be satisfied, in the presented approach  we restrict $x\in \mathcal{S}$. 

It is important to mention that few research work addressing the robust stabilization of time-delay systems via STA has been reported in the literature, see for instance \cite{WaaaahJeronimo2019, Caballero2018}. The reported results there are within the standard SMC framework, where one requires a transformation of the delay system to a regular form and the design of a sliding manifold.  Robust stabilization via the LR approach  requires neither  of both.

The note is organized as follows. The continuous LR methodology based on the STA is introduced in Section \ref{sec:STA}. The theoretical results are illustrated with an example  in Section \ref{sec:Example}, and   the paper ends with some final remarks in Section \ref{sec:conc}.

The following notation is adopted  throughout the paper.  The Euclidian norm for vectors and matrices  is denoted by $\|\cdot\|$.  The space of  continuous functions defined on $[-h,0]$ with values in $\mathbb{R}^n$ is denoted by $C\left([-h,0],\mathbb{R}^n\right)$  and it is equipped with the supremum norm
$$\|\varphi\|_{\mathcal{H}}:=\sup_{\theta \in [-h,0]}\|\varphi(\theta)\|.$$  The notation $x_t$ represents the state function $x_t(\theta)=x(t+\theta)$, $\theta\in [-h_m,0]$.
The space of positive real numbers is represented by $\mathbb{R}_+$.

\section{Continuous Lyapunov redesign for TDS}
\label{sec:STA}

We present the continuous LR methodology based on the STA. Let $w$ in \eqref{ec:sliding_var} be defined as a sliding variable. We address the robust stabilization of system \eqref{ec:linear_sys} by enforcing $w$ to be zero in finite time via a second order sliding mode controller and ensuring  asymptotic convergence of the system's solutions to the origin  on the sliding manifold $\mathcal{S}$.

Let us assume 
\begin{assum}
	\label{ass:matrix_B}
	$\rank B=k$.
\end{assum}
Consider the control law in system \eqref{ec:linear_sys} as
\begin{equation}
\label{ec:control}
u(t)=v_{nom}(t)+v(t),
\end{equation}
where
\begin{equation}
\label{ec:control_unc_sta}
v(t)=-\left(2B^TPB\right)^{-1}(2B^TP\left(\sum_{j=0}^{m}G_jx(t-h_j)\right)-u_{sta}(t)),
\end{equation}
with $P\in \mathbb{R}^{n\times n}$ from functional \eqref{ec:LK_functional_gen}, $u_{sta}$ denotes the STA  of variable gains introduced in \cite{Moreno2011,Vidaletal2016}  given by
\begin{equation}
\label{ec:STA}
\begin{split}
u_{sta}(t)=&-k_1(t,\bar x,\bar{\bar {x}} )\xi_1(w)+\rho(t)\\
\dot \rho(t)=&-k_2(t,\bar x,\bar{\bar {x}} )\xi_2(w),
\end{split}
\end{equation}
where $k_1$, $k_2$ and $\bar {\bar x}$ are specified later  on, $k_3$ is any positive real number and
\begin{equation*}
\begin{split}
\xi_1(w)&:=\dfrac{w(t)}{\|w(t)\|^{1/2}}+k_3w(t),\\
\xi_2(w)&:=\dfrac{w(t)}{2\|w(t)\|}+\frac{3k_3}{2}\dfrac{w(t)}{\|w(t)\|^{1/2}}+k_3^2w(t).
\end{split}
\end{equation*}

By setting $k_3=0$ and lifting the dependence of gains $k_1$ and $k_2$ on the state and delays, one recovers the standard STA, which has been recently studied for a class of time-delay systems in \cite{WaaaahJeronimo2019,Caballero2018}. {A remarkable} difference is that here the sliding manifold is not designed but results from the Lyapunov-Krasovskii functional used for the stability analysis of the nominal system. 

It is well-known that since the STA produces a continuous signal, it  cannot compensate  perturbations  satisfying $\|\delta(t,\bar x)\|\leq \rho_{\delta}(t,\bar x)$ \cite{Levant1993}.  That is the reason why we  introduce the following assumptions on the system uncertainty:
\begin{assum}
	\label{ass:delta}
	The uncertainty term can be divided as 
	\begin{equation}
	2B^TPB\delta(t,\bar x)=d_1(t,x)+\delta_z(t,\bar x),
	\end{equation}
	where  $d_1(t,x)=0$ if $x\in \mathcal{S}$  and $\delta_z$ is such that $$\frac{\partial \delta_z}{\partial  x(t-h_j)}B=0,\:j=0,\ldots,m.$$
\end{assum}
\begin{assum}
	\label{ass:bounds}
	There exist  known functions $\rho_1:\mathbb{R}_+\times \mathbb{R}^n\rightarrow \mathbb{R}_+$ and  $\rho_2:\mathbb{R}_+\times \mathbb{R}^{(m+1)n}\times \mathbb{R}\rightarrow \mathbb{R}_+$ such that 
	\begin{equation*}
	\begin{split}
	\|d_1(t,x)\|\leq& \rho_1(t,x)\|\xi_1(w)\|,\\
	\|	d_2(t,\bar x,\bar {\bar{ x}})\|\leq& \rho_2(t,\bar x,\bar {\bar{ x}})\|\xi_2(w)\|,
	\end{split}
	\end{equation*}
	where 
	\begin{equation*}
	d_2(t,\bar x,\bar {\bar{ x}}):=\dot \delta_z(t,\bar x)=\frac{\partial \delta_z}{\partial t}+\bar {\bar x}(t),
	\end{equation*}
	where $\bar {\bar x}(t):=\frac{\partial \delta_z}{\partial \bar x}\dot {\bar {x}}(t).$
\end{assum}

\begin{obs}
	Let $H_h$ denote the set of all possible sums of pairs of  delays $(h_i,h_j)$,  $i,j=0,\ldots,m$,  the vector $\bar{\bar{x}}$ depends on terms of the form  $x(t-h_{\kappa})$ with $h_\kappa\in H_h$. For example, for two delays $h_1$ and $h_2$, $H_h=\left\lbrace h_0, h_1,h_2,2h_1,2h_2,h_1+h_2\right\rbrace$ and $\bar {\bar x}$  depends on 
		$x(t),x(t-h_1),x(t-h_2),x(t-2h_1),x(t-2h_2),x(t-h_1-h_2).$

\end{obs}	

It is important to mention that whereas the component $d_1$ contains terms that vanish while $x\in \mathcal{S}$ for $t\geq T>0$, the component $d_2$ contains non-vanishing terms including those that are delay dependent, {i.e.} terms of the form $x(t-h_i)$, $i=1,\ldots,m$. These terms do not vanish until $t\geq T+h_{m}$. 

Let us consider the following Lyapunov function,  used in the proof of  Proposition \ref{prop:STA} in the Appendix (cf. with \cite{Moreno2011, Vidaletal2016}),
\begin{equation*}
V_{st}(w,z)=\gamma^TP_{st}\gamma,
\end{equation*}
where
\begin{equation*}
\gamma:=\begin{pmatrix}
\xi_1(w)\\ z
\end{pmatrix},\:P_{st}:=\begin{pmatrix}
\left(\beta +4\epsilon^2\right) I & -2\epsilon I \\
-2\epsilon I & I
\end{pmatrix},
\end{equation*}
and set the gains of the STA in \eqref{ec:STA} as
\begin{equation}
\label{ec:gains}
\begin{split}
&k_1(t,\bar x,\bar {\bar{x}})=\delta+\frac{1}{\beta}\left(\frac{1}{4\epsilon}\left(4\epsilon \rho_1+\rho_2\right)^2+2\epsilon \rho_2 +\epsilon+(2\epsilon+\rho_1)\left(\beta+4\epsilon^2\right)\right),\\
&k_2(t,\bar x,\bar {\bar{x}})=\beta+4\epsilon^2+2\epsilon k_1(t,\bar x,\bar {\bar{x}}),
\end{split}
\end{equation}
where $\delta,\: \beta,\:\epsilon \in \mathbb{R}_+$. In the next theorem we state the main result of the paper,  i.e. the robust stabilization of system \eqref{ec:linear_sys} by control \eqref{ec:control}.
\begin{theo}
	Suppose Assumptions \ref{ass:func}, \ref{ass:matrix_B}, \ref{ass:delta} and \ref{ass:bounds} are satisfied. Trajectories of closed-loop system \eqref{ec:linear_sys}, \eqref{ec:control}, with $v_{nom}$ and $v$ given by \eqref{ec:control_nom} and \eqref{ec:control_unc_sta} respectively, reach the sliding manifold $\mathcal{S}$  in finite time 
	\begin{equation*}
	T=\dfrac{2}{\eta_2}\ln \left(1+\dfrac{\eta_2}{\eta_1}\sqrt{V_{st}(w_0,z_0)}\right),
	\end{equation*}
	with
	\begin{equation*}
	\eta_1=\frac{\epsilon \sqrt{\lambda_{\min} (P_{st})}}{\lambda_{\max}(P_{st})},\:\eta_2=\frac{2\epsilon k_3}{\lambda_{\max}(P_{st})},
	\end{equation*}
	and after that they asymptotically  converge to the origin.
\end{theo}
\begin{proof}
The proof is splitted into two parts. First, it is proved  that with robust controller \eqref{ec:control} the trajectories of system \eqref{ec:linear_sys}  converge to the sliding manifold $\mathcal{S}$ in finite time, and then  that they asymptotically converge to the origin. 

 From  Assumption \ref{ass:delta} it follows that 
\begin{equation*}
\delta(t,\bar x)=\frac{1}{2}\left(B^TPB\right)^{-1}\left(d_1(t,x)+\delta_z(t,\bar x)\right),
\end{equation*}
hence, differentiating $w$ along the solutions of system \eqref{ec:linear_sys}, \eqref{ec:control},  yields
\begin{equation}
\label{ec:derivative_wsta}
\dot w(t)=u_{sta}(t)+d_1(t,x)+\delta_z(t,\bar x).
\end{equation}
By taking the change of variable $z(t):=\rho(t)+\delta_z(t,\bar x)$, where $\rho$ is from the STA \eqref{ec:STA}, we arrive at
\begin{equation}
\label{ec:sys_surf}
\begin{split}
\dot w(t)=&-k_1(t,\bar x,\bar {\bar x})\xi_1(w)+z(t)+d_1(t,x),\\
\dot z(t)=&-k_2(t,\bar x,\bar {\bar x})\xi_2(w)+d_2(t,\bar x,\bar {\bar{x}}).
\end{split}
\end{equation}
By Proposition \ref{prop:STA} in Appendix, system \eqref{ec:sys_surf} with gains given by \eqref{ec:gains}  is finite time stable.

Now, notice that closed-loop system \eqref{ec:linear_sys}, \eqref{ec:control}  is 
\begin{equation*}
\begin{split}
\dot x(t)=E\sum_{j=0}^{m} G_j x(t-h_j)+\frac{1}{2}B(B^TPB)^{-1}u_{sta}+B\delta(t,\bar x),
\end{split}
\end{equation*}
where $E=I-B(B^TPB)^{-1}B^TP$. It follows from  equation \eqref{ec:derivative_wsta} that
\begin{equation*}
\begin{split}
\dot x(t)=E\sum_{j=0}^{m} G_j x(t-h_j)+\frac{1}{2}B(B^TPB)^{-1}\dot w(t).
\end{split}
\end{equation*}
Thus, on the sliding manifold $\mathcal{S}$, 
\begin{equation}
\label{ec:reduced_system}
\begin{split}
\dot x(t)=E\sum_{j=0}^{m} G_j x(t-h_j),\:\:x\in \mathcal{S}.
\end{split}
\end{equation}
In order to prove that the trajectories asymptotically converge to the origin, let us consider a Lyapunov-Krasovskii functional $V$ of the form \eqref{ec:LK_functional_gen} satisfying conditions of Assumption \ref{ass:func}.  	The derivative of $V$ along the solutions of system \eqref{ec:reduced_system} satisfies
	\begin{equation*}
	\begin{split}
	\left.\dot V(x_t)\right\rvert_{\eqref{ec:reduced_system}}&\leq -\alpha_3\|x(t)\|^2-2x^T(t)PB(B^TPB)^{-1}B^TPE\sum_{j=0}^{m} G_j x(t-h_j)\\
	&=-\alpha_3\|x(t)\|^2-w^T(t)(B^TPB)^{-1}B^TPE\sum_{j=0}^{m} G_j x(t-h_j)\\
	&=-\alpha_3\|x(t)\|^2,\:x\in \mathcal{S}.
	\end{split}
	\end{equation*}
	This proves the asymptotic stability of system \eqref{ec:reduced_system}, and in turn completes the proof.
\end{proof}

The methodology for the design of robust controller   is summarized as follows: 
\begin{enumerate}
\item Propose a Lyapunov-Krasovskii functional of the form \eqref{ec:LK_functional_gen} that  satisfy the conditions of Assumption \ref{ass:func}. 
\item Select the \textit{sliding variable} $w(t)=2B^TPx(t)$, where matrix $P$ is from the Lyapunov-Krasovskii functional proposed in Step 1. 
\item Find $\rho_1$ and $\rho_2$ satisfying Assumption \ref{ass:bounds}.
\item Select controller $v(t)$ as  \eqref{ec:control_unc_sta} with a given sliding variable $w(t)$ in Step 2.
\item Fix the gains as in \eqref{ec:gains}  with upper-bounds given in Step 3.
\end{enumerate}


\section{Example}
\label{sec:Example}

The obtained results are illustrated with an example. 
The performance of  the unit control \eqref{ec:LR_discon}, continuous approximation \eqref{ec:LR_approx} and the STA \eqref{ec:STA} {are compared.} It is worth mentioning that the considered system  cannot be {robustly} stabilized by the approach proposed in \cite{Wu2004,Wu2009}. 


\begin{example}
	We consider a system of the form \eqref{ec:linear_sys} with $h_1=h=2$ and matrices \cite{Gouaisbautetal2002}
	\begin{equation*}
	A_0=\begin{pmatrix}
	2 & 0\\
	1.75 & 0.25
	\end{pmatrix},\:A_1=\begin{pmatrix}
	-1 & 0\\
	-0.1 & -0.25
	\end{pmatrix},\:B=\begin{pmatrix}
	1\\1
	\end{pmatrix}.
	\end{equation*}
	For the sake of simulation, consider the uncertainty term as $$\delta(t,x)=\dfrac{1}{3}\left(\sin(t)+2(x_1(t-h)-x_2(t-h))\right).$$

	Notice that the associated nominal system is not  stabilizing by a memoryless feedback since the pair $(A_0,B)$ is not controllable. The robust stabilization of this system has been addressed in \cite{Gouaisbautetal2002}  via SMC. We next  present the simulation results obtained  with control
	\begin{equation*}
	u(t)=v_{nom}(t)+v(t),
	\end{equation*}
	where $v(t)$ is considered to be of three classes: unit control of the form \eqref{ec:LR_discon}, continuous control with  approximation \eqref{ec:LR_approx} with $\varepsilon=0.05$, and continuous control based on STA \eqref{ec:STA}.  In all of them we consider the same nominal control and the same associated Lyapunov-Krasovskii functional of the  form \eqref{ec:LK_functional_gen}:
	\begin{equation*}
	V(x_t)=x^T(t)Px(t)+\int_{-h}^{0}\left(x^T(t+\xi)Qx(t+\xi)d\xi+\int_{\theta-h}^{0}x^T(t+\xi)Rx(t+\xi)d\xi\right)d\theta
	\end{equation*}
	where
	\begin{align*}
	P&=\begin{pmatrix}
	 2.8063  &   0.8062\\
	0.8062   & 0.6559
	\end{pmatrix}, \: Q=\begin{pmatrix}
	9.1429 &   3.3997\\
	3.3997  &  1.3327
	\end{pmatrix},\\R&=\begin{pmatrix}
	4.6439 &   1.5632\\
	1.5632  &  0.6352
	\end{pmatrix}.
	\end{align*}
	Stabilizing gains of the nominal system are found to be
	\begin{equation*}
	K_1=\begin{pmatrix}
	-3.7648 & -0.73
	\end{pmatrix},\:K_2=\begin{pmatrix}
	1.1964 & 0.1723
	\end{pmatrix}.
	\end{equation*}
	The sliding variable is 
	\begin{equation*}
	w(t)=2B^TPx(t)=7.2251x_1(t)+2.9244x_2(t).
	\end{equation*}
	
	Let us define $c=\frac{2}{3}B^TPB=3.3832$ and consider the gain $k_3=0$.  The  uncertainty satisfies Assumption \ref{ass:delta} with $d_1(t,x)=0$ and 
	\begin{equation*}
	\delta_z(t,\bar x)=c\left(\sin(t)+2(x_1(t-h)-x_2(t-h))\right).
	\end{equation*}
	It is clear that $\dfrac{\partial \delta_z}{\partial x(t-h)}B=0$. Then, the uncertainty also satisfies Assumption \ref{ass:bounds} with
	$\rho_1(t,x)=0$ and, 
	\begin{equation*}
	\rho_2(t,\bar x,\bar {\bar x})=(4c^2\left(1+2|0.5x_2(t-2h)-1.8x_1(t-2h)|^2\right)+2c^2|x_1(t-h)-x_2(t-h)|^2)^{1/2}.
	\end{equation*}
	For unit control \eqref{ec:LR_discon} and continuous approximation \eqref{ec:LR_approx},  consider $\rho_{\delta}(t,\bar x)=\frac{1}{3}(1+2|x_1(t-h)-x_2(t-h)|)$.
	
	Set then the gains $k_1$ and $k_2$ as in \eqref{ec:gains} with the parameters
	\begin{equation*}
	\delta= 1.5,\:\beta= 1,\:\epsilon=0.3.
	\end{equation*}
	
	Figure \ref{fig:sol} depicts the norm of the closed-loop solution. One observes  that the  LR based on STA has a better performance in comparison with the continuous approximation  \eqref{ec:LR_approx} and the unit control \eqref{ec:LR_discon}. Indeed, closed-loop solution of the system with \eqref{ec:LR_approx} does not converge to the origin but it is restricted to a small neighborhood. Figure \ref{fig:control} shows the control laws, where the reduction of chattering with respect to the unit control is clearly visualized.

	\begin{figure}[htpb!]
		\centering
		\includegraphics[scale=1]{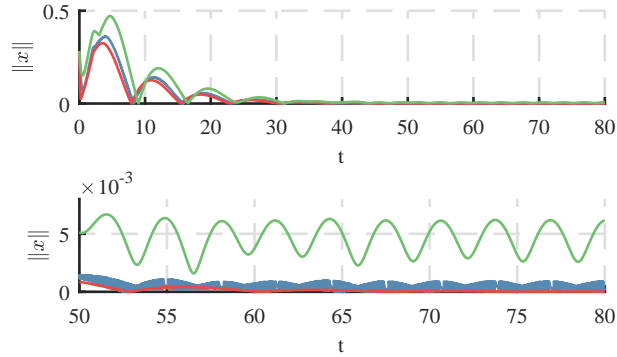}
		\caption{Top: Norm of the closed-loop solution obtained with  LR based on unit control (blue), continuous LR based on approximation \eqref{ec:LR_approx} (green), and  LR based on STA (red). Bottom: Simulation result of the last 30 seconds}
		\label{fig:sol}
	\end{figure}
	
	\begin{figure}[htpb!]
		\centering
		\includegraphics[scale=1]{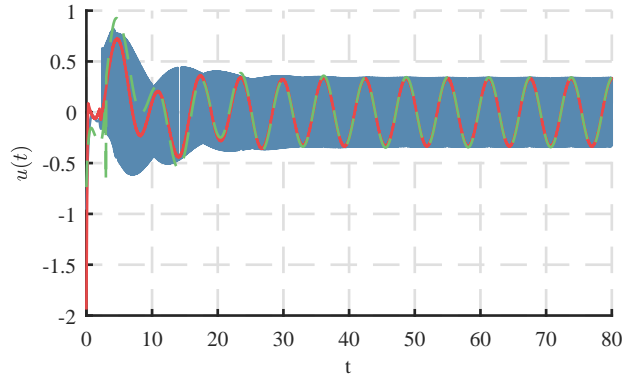}
		\caption{Top: Control signals of LR based on LR based on unit control (blue), continuous LR based on approximation \eqref{ec:LR_approx} (green), and  LR based on STA (red). Bottom: Simulation result of the last 30 seconds.}
		\label{fig:control}
	\end{figure}
	
\end{example}

\section{Conclusions}
\label{sec:conc}

The robust stabilization of a class of uncertain systems with delays via a  new continuous LR methodology based on the STA {was addressed.} A remarkable feature of the proposed approach is that it allows one to  ensure asymptotic stability of the system {using continuous control signals. } 

As a direct consequence of the considered class of Lyapunov-Krasovskii functionals, the associated sliding variable is delay-free. The latter allowed us to combine the STA with the LR technique without eliminating the sliding modes (see Chapter 13 in \cite{Perruquetti2002} {and Section IV in \cite{EfimovAleksandrov2020})}, providing a flexible, robust design methodology that can be easily extended to more complex scenarios studied in SMC for {delay-free}  systems.

\appendix

\section{Proposition \ref{prop:STA}}
\label{app:STA}
\begin{propo}
	\label{prop:STA}
	Under Assumptions \ref{ass:func}, \ref{ass:matrix_B}, \ref{ass:delta} and \ref{ass:bounds}, system 
	\begin{equation*}
	\begin{split}
	\dot w(t)=&-k_1(t,\bar x,\bar {\bar x})\xi_1(w)+z(t)+d_1(t,x),\\
	\dot z(t)=&-k_2(t,\bar x,\bar {\bar x})\xi_2(w)+d_2(t,\bar x,\bar {\bar{x}}),
	\end{split}
	\end{equation*}
	 with $k_1$ and $k_2$ given by \eqref{ec:gains},  is finite time stable  with 
	\begin{equation*}
	T=\dfrac{2}{\eta_2}\ln \left(1+\dfrac{\eta_2}{\eta_1}\sqrt{V_{st}(w_0,z_0)}\right). 
	\end{equation*}
\end{propo} 
\begin{proof}
	The proof  follows the same arguments as those presented in \cite{Vidaletal2016} up to the dependence of the gains on the delayed states.
	 Let us consider the Lyapunov function
	\begin{equation*}
	V_{st}(w,z)=\gamma^TP_{st}\gamma,
	\end{equation*}
	where
	\begin{equation*}
	\gamma=\gamma(w,z):=\begin{pmatrix}
	\xi_1(w)\\ z
	\end{pmatrix},\:P_{st}:=\begin{pmatrix}
	\left(\beta +4\epsilon^2\right) I & -2\epsilon I \\
	-2\epsilon I & I
	\end{pmatrix}.
	\end{equation*}
	The Lyapunov function $V_{st}$ is differentiable everywhere except on the set $\mathcal{W}:=\{(w,z)\in \mathbb{R}^{2k}: w=0 \}$. Let us  define $\xi_1'(w):=\dfrac{d}{dw}\xi_1(w)\in \mathbb{R}^{k\times k}$. We observe that
	\begin{equation*}
	\begin{split}
	\xi_1'(w)=&\|w\|^{-1/2}I-\dfrac{1}{2}ww^T\|w\|^{-5/2}+k_3I\\
	=&\frac{1}{\|w\|^{1/2}}\left(I-\frac{1}{2}\frac{ww^T}{\|w\|^2}\right)+k_3I
	\end{split}
	\end{equation*}
	for any  different from zero $w\in \mathbb{R}^k$.  We recall from \cite{Vidaletal2016}  some properties of that are useful throughout the proof:
	\begin{enumerate}
		\item $\xi_2(w)=\xi_1'(w)\xi_1(w),\:\:\forall w\neq 0$.
		\item Matrix $\xi_1'(w)$ is symmetric and positive definite for any different from zero $w\in \mathbb{R}^k$. Moreover,
		\begin{equation*}		
		\lambda_{\min}(\xi_1'(w))\|y\|^2\leq y^T\xi_1'(w)y,\:\:\forall y\in \mathbb{R}^k,
		\end{equation*}
		with $\lambda_{\min}(\xi_1'(w))=\dfrac{1}{2\|w\|^{1/2}}+k_3$.
		\item $\|\xi_1'(w)\|=\dfrac{1}{\|w\|^{1/2}}+k_3.$ 
	\end{enumerate}
	By Property 1, 
	\begin{equation*}
	\begin{split}
	\dot \gamma =\begin{pmatrix}
	\xi_1'(w) \dot w\\
	\dot z
	\end{pmatrix}&=\begin{pmatrix}
	\xi_1'(w)\left(-k_1\xi_1(w)+z+d_1\right)\\
	-k_2\xi_1'(w)\xi_1(w)+d_2
	\end{pmatrix}\\
	&=A_{st}\gamma +f,\: (w,z)\in \mathbb{R}^{2k}\setminus \mathcal{W},
	\end{split}
	\end{equation*} 
	where
	\begin{equation*}
	A_{st}=A_{st}(t,\bar x,\bar {\bar x}):=\begin{pmatrix}
	-k_1(t,\bar x,\bar {\bar x})\xi_1'(w) & \xi_1'(w)\\
	-k_2(t,\bar x,\bar {\bar x})\xi_1'(w) & 0
	\end{pmatrix} 
	\end{equation*}
	and
	\begin{equation*}
	f=f(t,\bar x,\bar {\bar x}):=\begin{pmatrix}
	d_1(t,x)\\
	d_2(t,\bar x,\bar {\bar x}))
	\end{pmatrix}.
	\end{equation*}
	Hence,
	\begin{equation*}
	\dfrac{d}{dt}V_{st}(w,z)=-\gamma^TQ\gamma+2\gamma^TP_{st}f.
	\end{equation*}
	where
	\begin{align*}
	Q_{st}&:=-\left(P_{st}A_{st}+ A_{st}^TP_{st}\right)\\
	&=\begin{pmatrix}
	2\beta k_1-4\epsilon \left( \beta+4\epsilon^2\right)\xi_1'(w) & 0\\
	0 & 4\epsilon \xi_1'(w)
	\end{pmatrix}.
	\end{align*}
	From Assumption \ref{ass:bounds},  it follows  that
	\begin{equation*}
	\begin{split}
	\gamma^TP_{st}f\leq& (\beta+4\epsilon^2)\|\xi_1^T\xi_1'(w)\|\|d_1\|+2\epsilon \|z\|\|\xi_1'(w)\|\|d_1\|+\left(2\epsilon \|\xi_1\|+\|z\|\right)\|d_2\|\\
	\leq& (\beta+4\epsilon^2)\|\xi_1^T\xi_1'(w)\|\rho_1\|\xi_1\|+2\epsilon \|z\|\|\xi_1'(w)\|\rho_1\|\xi_1\|+\left(2\epsilon \|\xi_1\|+\|z\|\right)\rho_2\|\xi_2\|.
	\end{split}
	\end{equation*}
	Then,  from  equality  $\|\xi_1^T \xi_1'\|=\lambda_{\min}(\xi_1'(w)) \|\xi_1\|$  and $2\|\xi_1'(w)\|\leq 4\lambda_{\min}(\xi_1'(w))$, we obtain
	\begin{equation}
	\label{ec:pert_term}
	2\gamma^TP_{st}f\leq 2\lambda_{\min}(\xi_1'(w))(   (\beta+4\epsilon^2)\|\xi_1\|^2\rho_1+4\epsilon\|z\|\|\xi_1\|\rho_1+2\epsilon \|\xi_1\|^2\rho_2+\|z\|\|\xi_1\|\rho_2),
	\end{equation}
	and  from Property 2  we have that for any $\gamma \in \mathbb{R}^{2k}$
	\begin{equation}
	\label{ec:nom_term}
	-\gamma^T Q_{st}\gamma\leq -\lambda_{\min}(\xi_1'(w))\left(\left(2\beta k_1-4\epsilon \left(\beta+4\epsilon^2\right)\right)\|\xi_1\|^2+4\epsilon \|z\|^2\right). 
	\end{equation}
	By \eqref{ec:nom_term} and \eqref{ec:pert_term}, 
	\begin{equation*}
	\dfrac{d}{dt}V_{st}(w,z)\leq -\lambda_{\min}(\xi_1'(w))\hat \gamma^T \hat Q \hat \gamma,
	\end{equation*}
	with $\hat \gamma:=\begin{pmatrix}
	\|\xi_1\| & \|z\|
	\end{pmatrix}^T$ and
	\begin{equation*}
	\hat Q=\begin{pmatrix}
	2\beta k_1-(4\epsilon+2\rho_1)(\beta+4\epsilon^2)-4\epsilon \rho_2 & -4\epsilon \rho_1-\rho_2\\
	\star & 4\epsilon
	\end{pmatrix}.
	\end{equation*}
	Considering $k_1$ as in \eqref{ec:gains}
	one has that  $\hat Q-2\epsilon I>0$, hence
	\begin{equation*}
	\dfrac{d}{dt}V_{st}(w,z)\leq -2\epsilon \lambda_{\min}(\xi'(w))\|\hat \gamma\|^2.
	\end{equation*}
	Since $\|w\|^{1/2}\leq \|\xi_1\|\leq \|\hat \gamma\|$ and $\lambda_{\min}(P_{st})\|\hat \gamma\|^2\leq V_{st}(w,z)\leq \lambda_{\max}(P_{st})\|\hat \gamma\|^2$, 
	\begin{equation*}
	\begin{split}
	\dfrac{d}{dt}V_{st}(w,z)&\leq -2\epsilon \left(\frac{1}{2\|w\|^{1/2}}+k_3\right)\|\hat \gamma\|^2\\
	&\leq -\eta_1 \sqrt{V_{st}(w,z)}-\eta_2 V_{st}(w,z),
	\end{split}
	\end{equation*}
	where
	\begin{equation*}
	\eta_1=\frac{\epsilon \sqrt{\lambda_{\min} (P_{st})}}{\lambda_{\max}(P_{st})},\:\eta_2=\frac{2\epsilon k_3}{\lambda_{\max}(P_{st})}.
	\end{equation*}
	Finally, as the solution of the differential equation
	\begin{equation*}
	\dot v_c(t)=-\eta_1 \sqrt{v_c(t)}-\eta_2 v_c(t), \:v_c(0):=v_{c_0}
	\end{equation*}
	is determined by
	\begin{equation*}
	v_c(t)=e^{-\eta_2 t }\left(v_{c_0}^{1/2}+\dfrac{\eta_1}{\eta_2}\left(1-e^{\frac{{\eta_2}}{2}t}\right)\right)^2
	\end{equation*}
	it follows from the comparison theorem that $(w,z)$ converges to zero in finite time
	\begin{equation*}
	T=\dfrac{2}{\eta_2}\ln \left(1+\dfrac{\eta_2}{\eta_1}\sqrt{V_{st}(w_0,z_0)}\right).
	\end{equation*}	
\end{proof}
%
%
%
%

\bibliographystyle{IEEEtran}        
\bibliography{referencesTAC}

\end{document}